\documentclass[12pt]{iopart}
\usepackage{graphicx}
\usepackage{setspace}
\usepackage{texdraw}
\usepackage{amsfonts}

\newcommand{\be}{\begin{equation}}
\newcommand{\ee}{\end{equation}}
\newcommand{\bea}{\begin{eqnarray}}
\newcommand{\eea}{\end{eqnarray}}

\begin{document}

\title{Temperature Correlation   of Quantum Spins}

\author{ \ A. R. Its\dag , \ V. E. Korepin \maltese , \ A.G. Izergin $\bullet$,  \ N.A.
Slavnov $\ddagger$ }

\address{ \dag\ Department of Mathematical Sciences, Indiana
University-Purdue University Indianapolis, Indianapolis, IN
46202-3216, USA}
\address{$\maltese$ \ C.N.\ Yang Institute for Theoretical Physics, State
University of New York at Stony Brook, Stony Brook, NY 11794-3840,
USA}
\address{$\bullet$  St. Petersburg Department 
of Steklov Mathematical Institute, RAS 27, Fontanka,  191023,  St. Petersburg,  Russia   }
\address{$\ddagger$ \ Steklov Mathematical Institute,
Gubkina 8, Moscow, 119991, Russia } 

\begin{abstract}
We consider  isotropic version of XY model in  transverse magnetic field [in one space
dimension].    We calculated  of asymptotic of temperature correlations.
  We represent the quantum correlation function as a tau function of a classical
completely
integrable
differential equation.  This is the well-known Ablowitz-Ladik lattice equation.
\end{abstract}
\vfill
\vfill
\section{Introduction}
The XY model was introduced and studied by E. Lieb, T. Schultz and D.
Mattis
[1].  It
describes the interaction of spins 1/2 situated on a 1-dimensional periodic
lattice.  The
Hamiltonian of the model is

$$ H =- \sum_n\left[\sigma^x_n \sigma^x_{n+1} + \sigma^y_n\sigma^y_{n+1} +
h\sigma^z_n\right]. \eqno (1)$$
Here $\sigma$ are Pauli matrices, $h$ is transverse magnetic field and $n$
enumerates the
sites of the lattice.  At zero temperature the problem of evaluation of
asymptotics of
correlation functions was solved in [2,3].  Here we consider the
temperature correlation function.
$$ g(n,t) = \frac{Tr
\{e^{-\frac{H}{T}}\sigma^+_{n_2}(t_2)\sigma^-_{n_1}(t_1)\}}{Tr
e^{-\frac{H}{T}}}, \ \ \ \
n = n_2-n_1, \ \ \ \ t=t_2-t_1 \eqno (2) $$
for the infinite lattice.
We consider finite temperature $0 < T < \infty$ and a moderate magnetic
field
$0 \leq h <
2$.

We evaluated the asymptotics in cases where both space and time separation
go
to infinity $n
\rightarrow \infty$, $t\rightarrow \infty$, in some direction $\varphi$

$$ \frac{n}{4t} = \cot \varphi, \ \ \ \ 0 \leq \varphi \leq \frac{\pi}{2}.
\eqno (3)$$
In accordance with our calculations, correlation function $g(n,t)$ decays
exponentially in any direction, but the rate of
decay depends on the direction.  In the space like direction, $0 \leq
\varphi <
\frac{\pi}{4}$, the asymptotics are

$$ g(n,t) \rightarrow C \exp \left\{ \frac{n}{2\pi}\int^\pi_{-\pi}dp\ln
\left|\tanh
(\frac{h-2\cos p}{T})\right|\right\}. \eqno (4)$$
In the time like direction $\frac{\pi}{4} < \varphi \leq \frac{\pi}{2}$,
the
asymptotics are
different:

$$ g(n,t) \rightarrow C t^{(2\nu^2_++2\nu^2_-)}\exp \left\{
\frac{1}{2\pi} \int^\pi_{-\pi}dp
\left|n-4t\sin p\right|\ln\left|\tanh(\frac{h-2\cos p}{T})\right|\right\}.
\eqno (5)$$
The values $\nu_\pm$, which define the pre-exponent, are

$$ \nu_+ = \frac{1}{2\pi}\ln\left|\tanh \left( \frac{h-2\cos
p_0}{T}\right)\right|$$

$$ \nu_- = \frac{1}{2\pi}\ln\left|\tanh \left(
\frac{h+2\cos\rho_0}{T}\right)\right|\eqno (6)$$
where $\frac{n}{4t} = \sin p_0$.  Equation (5) is valid in the whole time
like cone,
with exception of one direction $h=2\cos p_0$.  Higher asymptotic
corrections will
modify formulae by a factor of $(1+c(t,x))$  ($c$ decays exponentially in the
space-like
region and as $t^{-1/2}$ in the time-like region).  Also, it should be
mentioned that the
constant factor $C$ in (4) does not depend on the direction $\varphi$, but
does depend on
$\varphi$ in (5).
We want to emphasize that for the pure time direction, $\varphi = \pi/2$,
the leading
factor in the
asymptotics (exponent in (5)) was first obtained in [10].

To
derive these formulae we went through a few steps.

The first step: The explicit expression for eigenfunctions of the
Hamiltonian (1) (see [1])
was used to represent the correlation function as a determinant of an
integral
operator (of Fredholm type) [8].  In order to explain we need to introduce
some
notation.
Let us consider the integral operator $\hat V$.  Its kernel is equal to

$$ V(\lambda\mu) = \frac{e_+(\lambda)e_-(\mu) -
e_-(\lambda)e_+(\mu)}{\pi(\lambda -
\mu)}. \eqno (8)$$
Here $\lambda$ and $\mu$ are complex variables, which go along the  circle
$\left|\lambda\right| = \left|\mu\right| = 1$ in the positive direction.
The functions $e_\pm$ are
$$ e_-(\lambda) = \lambda^{-n/2} \cdot e^{-it(\lambda+1/\lambda)}
\sqrt{v(\lambda)}, \eqno
(9)$$
where

$$ v(\lambda) = \left\{ 1 + \exp \left[ \frac{2h-2(\lambda +
\frac{1}{\lambda})}{T}\right]\right\}^{-1}, \eqno (10)$$
and

$$ e_+(\lambda) = e_-(\lambda)E(x,t,\lambda). \eqno (11)$$
Here $E$ is defined as an integral

$$ E(n,t,\lambda) = \frac{1}{\pi}v.p.\int\exp\{ 2it(\mu + \frac{1}{\mu})\}
\cdot
\frac{\mu^nd\mu}{\mu-\lambda}. \eqno (12)$$
It is convenient to define functions $f_\pm(\lambda)$ as solutions of the
following
integral equations:

$$ (I + \hat V)f_k = e_k. \eqno (13)$$
Here $I$ is the identity operator and $k =\pm$.  Next we define the
potentials
$B_{kj} (k,j
= \pm)$:

$$ B_{kj}(n,t) = \frac{1}{2\pi i} \int f_k(\lambda)
e_j(\lambda)\frac{d\lambda}{\lambda}.
\eqno (14)$$
They depend on space and time variables $n,t$.  These we shall use to
define new
potentials $b_{kj}$:
$$ \begin{array}{ll} b_{--}(n,t) = B_{--}(n,t), \\ \\

 b_{++}(n,t) = B_{++}(n,t) - 2iG(n,t)B_{+-}(n,t) - G(n,t). \end{array}
\eqno (15)$$

\noindent Here we used the function

$$ G(n,t) = \frac{1}{2\pi i}\int \lambda^{n-1}\exp \{ 2it(\lambda +
\frac{1}{\lambda})\}
d\lambda. \eqno (16)$$
Now all the notation is ready to write a determinant formula
for the correlation function $g(n,t)$ (see (4)):

$$ g(n,t) = e^{-2iht}b_{++}(n,t)\exp \{ \sigma(n,t)\}. \eqno (17)$$
Here $e^\sigma$ is a determinant of the integral operator

$$ \exp\{ \sigma(n,t)\} = \det(1+\hat V). \eqno (18)$$

Second step: Formulae (8)-(15) can be used to show that the potentials
$b_{++}$
and $b_{--}$
satisfy a system of nonlinear differential equations.
$$ \frac{i}{2} \frac{\partial}{\partial t} b_{--}(n,t) = (1 +
4b_{--}(n,t)b_{++}(n,t))(b_{--}(n+1,t) + b_{--}(n-1,t)) $$

$$ -\frac{i}{2} \frac{\partial}{\partial t} b_{++}(n,t) =
(1+4b_{--}(n,t)b_{++}(n,t))(b_{++}(n+1,t) + b_{++}(n-1,t)). \eqno (19)$$
The derivation of these equations is similar to [4,5,7].  Equations (19)
are
completely
integrable differential equations.  They were first discovered by Ablowitz
and Ladik [9]
as an integrable discretization of the nonlinear Schroedinger equation.  The
logarithmic
derivatives
of $\sigma(x,t)$ (see (18)) can be expressed in terms of solutions of the
system (19):

$$ \begin{array}{ll}
\frac{\partial^2\sigma(n,t)}{16\partial t^2} & = 2b_{--}(n,t)b_{++}(n,t) -
b_{++}(n-1,t)b_{--}(n+1,t) - \\ \\

& -b_{--}(n-1,t)b_{++}(n+1,t) -
4b_{++}(n,t)b_{--}(n,t)[b_{++}(n-1,t)b_{--}(n+1,t) +\\
\\

& b_{--}(n-1,t)b_{++}(n+1,t)] \end{array} \eqno (20)$$

$$ \sigma(n+1,t) + \sigma(n-1,t) -2\sigma(n,t) =
\ln[1+4b_{--}(n,t)b_{++}(n,t)] \eqno
(21)$$

$$ \frac{\partial}{\partial t}[\sigma(n+1,t) - \sigma(n,t)] =
8i[b_{++}(n+1,t)b_{--}(n,t) - b_{++}(n,t)b_{--}(n+1,t)]. \eqno (22)$$
This shows that the quantum correlation function $g$ (2) can be expressed
in
terms of the
solution of the system (19).
The meaning of all these formulae is that the correlation function of the
$XY$ model is
the $\tau$-
function (in a sense of the well-known works [11,12]) of Ablowitz-Ladik's
differential-difference equations.  In the papers [5,6,7] the relation
between the
$\tau$
functions of the classical partial differential equations and quantum
correlation
functions, together with the history of the question, is explained in more
detail.  It
is also worth mentioning that the idea to connect quantum correlation
functions and
classical completely integrable systems goes back to the work [13] and was
first applied
to the $XY$ model in [14].

Third step: In order to evaluate the asymptotics one should solve
Ablowitz-Ladik's
differential equation.  Initial data can be extracted from the integral
representations
(8)-(18).  We use the Riemann-Hilbert problem in order to evaluate the
asymptotics
of the solution
of equation (19).  It is quite similar to the nonlinear Shrodinger  case
[6,7].

Finally, let us explain the physical meaning of our asymptotic formula
(4).  We start
from the expression for the free energy [1]:

$$ f(h)=-h-\frac{T}{2\pi}\int^\pi_{-\pi}dp \ln\left(1 + \exp[ \frac{4\cos
p-2h}{T}]\right).  \eqno (23)$$
We emphasize the dependence on the magnetic field $h$.  The definition of
$f(h)$ is
standard:
$$ Tre^{-\frac{H}{T}} = \exp\{ -\frac{L}{T}f(h)\}. \eqno (24)$$
Here $L$ is the length of the box.  Let us use Jordan-Wigner transformation
to transform
correlator (2) (in the equal time case):

$$ \sigma^+_{n_2}(0)\sigma^-_{n_1}(0) = \psi_{n_2}\exp \left\{ i\pi
\sum^{n_2-1}_{k=n_1+1}\psi^+_k\psi_k\right\}\psi^+_{n_1}, \ \ \ \
\psi^+_k\psi_k=\frac{1}{2}(1-
\sigma^z_k). \eqno (25)$$
Here $\psi_k$ is a canonical Fermi field.  We note that numerator in
(2) differs
from the denominator by replacement of the magnetic field $h\rightarrow
h-i\pi T/2$
on the space
interval $[n_1+1,n_2-1]$.  This leads us to the following asymptotic
expression for
correlator $g(n,0)$

$$ g(n,0) \rightarrow \exp\left\{ Re \frac{n}{T} \left[ f(h)-f(h-
\frac{i\pi T}{2})\right]\right\}. \eqno (26) $$
The reason we wrote $Re$ is that $i\pi$ in (25) can be replaced by
$-i\pi$.
It is remarkable that (26) coincides with the correct answer (4).  It is
also worth
mentioning that to go to the exponent in (5) one should replace the
differential
$d(np)$ by the
expression $|d(np-t\varepsilon(p)|$, where $\varepsilon(p) =-4\cos p+2h$ is
the energy of
the quasiparticle of the model.
\vskip .2in
\noindent{\Large Acknowledgements}

This work was partially supported by NSF Grant No. PHY-9107261
One of the authors (A.G.I.) is grateful to the Laboratoire de
Physique
Theorique in Ecole Normale Superrieure de Lyon (France) for their warm
hospitality.


\vskip 1cm
\begin{thebibliography}{5}
\bibitem{1} E. Lieb, T. Schultz and D. Mattis, Ann. Phys. (N.Y.) {\bf 16},
406 (1961).

\bibitem{2} B.M. McCoy, J.H.H. Perk, R.E. Shrock, Nucl. Phys. B {\bf 220},
269 (1983);
Nucl. Phys. B {\bf 220}, 35 (1983).

\bibitem{3} H.G. Vaidya, C.A. Tracy, Phys. Lett. A {\bf 68}, 378 (1978).

\bibitem{4} A.R. Its, A.G. Izergin, V.E. Korepin, N.A. Slavnov, J. Mod.
Phys. B. {\bf
4}, 1003 (1990).

\bibitem{5} A.R. Its, A.G. Izergin, V.E. Korepin, N.A. Slavnov, in the book
``Important
Developments in Soliton Theory, 1980-1990", eds. A.S. Fokas, V.E. Zakharov,
Springer-Verlag, 1992.

\bibitem{6} A.R. Its, A.G. Izergin, V.E. Korepin, G.G. Varzugin, Physica D,
{\bf 54},
351 (1992).

\bibitem{7} V.E. Korepin, A.G. Izergin, N.M. Bogolinbov, ``Quantum Inverse
Scattering
Method, Correlation Functions and Algebraic Bethe Ansatz" Cambridge
University Press,
(1992).

\bibitem{8} F. Colomo, A.G. Izergin, V.E. Korepin, V. Tognetti,
``Determinant
Representation for Correlation Functions in the XXO Heisenberg Chain"
preprint
ITP-SB-92-12, accepted in Phys. Lett.

\bibitem{9} M.J. Ablowitz, J.F. Ladik, Stud. Appl. Math. {\bf 55}, 213
(1976).

\bibitem{10} P. Deift, X. Zhou, preprint, Courant Mathematical Institute
(1992).
\bibitem{11} M. Jimbo, T. Miwa, Y. Mori, and M. Sato, Physica {\bf 1D},
80-158 (1980).

\bibitem{12} M. Jimbo, T. Miwa, and K. Ueno, Physica {\bf 2D}, 306-352
(1981).

\bibitem{13} E. Barouch, B.M. McCoy, T.T. Wu, Phys. Rev. Lett., {\bf 31}, 1409
(1973).

\bibitem{14} J.H.H. Perk, Phys. Lett., {\bf 79A}, 1, 1 (1980).
\end{thebibliography}
\end{document}